\documentclass[twocolumn,aps,prl,superscriptaddress]{revtex4-1}
\usepackage{amsfonts}
\usepackage{amsmath}
\usepackage{txfonts}
\usepackage{amssymb}
\usepackage{bbold}
\usepackage{amsbsy} %for sigma in bold face
\usepackage{epsfig}
\usepackage{cancel}
\usepackage{float}
\usepackage{romannum}
\usepackage{color}
\usepackage[colorlinks=true,citecolor=blue,linkcolor=red]{hyperref}
\usepackage[multiple]{footmisc}

\def\bea{\begin{eqnarray}} \def\eea{\end{eqnarray}}

\def\bsigma{{\boldsymbol \sigma}}

\def\b1{{\bf 1}}

\def\bk{{\bf k}}

\def\bd{{\bf d}}

\def\bn{\hat{\bn}}

\def\sphere#1{S^{\!#1}}
\def\disc#1{D^{#1}}
\def\mcT{\mathcal{T}}
\def\mcP{\mathcal{P}}
\DeclareMathAlphabet{\pazocal}{OMS}{zplm}{m}{n}
\def\mcC{\pazocal{C}}
\def\abs#1{\left|{#1}\right|}

\usepackage[dvipsnames]{xcolor}
\definecolor{TB}{rgb}{1,0.5,0}
 \def\TB{\color{black}}

\begin{document}

\title{Conversion rules for Weyl points and nodal lines in topological media}

\author{Xiao-Qi Sun}
\affiliation{Department of Physics, McCullough Building, Stanford University, Stanford, California 94305-4045, USA}
\affiliation{Stanford Center for Topological Quantum Physics, Stanford University, Stanford, California 94305-4045, USA}
\author{Shou-Cheng Zhang}
\affiliation{Department of Physics, McCullough Building, Stanford University, Stanford, California 94305-4045, USA}
\affiliation{Stanford Center for Topological Quantum Physics, Stanford University, Stanford, California 94305-4045, USA}
\author{Tom{\'a}{\v s} Bzdu{\v s}ek}
\affiliation{Department of Physics, McCullough Building, Stanford University, Stanford, California 94305-4045, USA}
\affiliation{Stanford Center for Topological Quantum Physics, Stanford University, Stanford, California 94305-4045, USA}

\begin{abstract}
According to a widely-held paradigm, a pair of Weyl points with opposite chirality mutually annihilate when brought together. In contrast, we show that such a process is strictly forbidden for Weyl points related by a mirror symmetry, provided that an effective two-band description exists in terms of orbitals with opposite mirror eigenvalue. Instead, such a pair of Weyl points convert into a nodal loop inside a symmetric plane upon the collision. Similar constraints are identified for systems with multiple mirrors, facilitating previously unreported nodal-line and nodal-chain semimetals that exhibit both Fermi-arc and drumhead surface states. We further find that Weyl points in systems symmetric under a $\pi$-rotation composed with time-reversal are characterized by an additional integer charge that we call helicity. A pair of Weyl points with opposite chirality can annihilate only if their helicities also cancel out. We base our predictions on topological crystalline invariants derived from relative homotopy theory, and we test our predictions on simple tight-binding models. The outlined homotopy description can be directly generalized to systems with multiple bands and other choices of symmetry.
\end{abstract}

\maketitle

\emph{Introduction}.--- The study of topological insulators and superconductors~\cite{Thouless:1982,Volovik:2003,Kane:2005,Qi:2008,Schnyder:2008,Kitaev:2009,Ryu:2010,Zak:1989,King-Smith:1993,Hasan:2010,Qi:2011,Soluyanov:2011,Bernevig:2013,Slager:2013,Chiu:2016} has recently broadened into a fruitful research of \emph{gapless} topological phases \TB{with band-structure nodes (BSNs) near the Fermi energy, including} nodal-point~\cite{Murakami:2007,Wan:2011,Weng:2015,Xu:2015a,Lv:2015,Soluyanov:2015,Wang:2016b,Young:2012,Wang:2012,Neupane:2014,Liu:2014a,Liu:2014b,Chen:2015b,Witten:2015} and nodal-line~\cite{Burkov:2011,Heikkila:2015b,Bian:2015a,Bian:2015b,Chen:2015,Schoop:2016,Fang:2016,Hirayama:2017} semimetals. Prominent BSNs include \emph{Weyl points} (WPs)~\cite{Murakami:2007,Wan:2011,Weng:2015,Xu:2015a,Lv:2015,Soluyanov:2015,Wang:2016b} stabilized by the Chern number (also called \emph{chirality}), which occur generically in systems without $\mcP\mcT$-symmetry~\cite{Hosur:2013}. By bulk-boundary correspondence, surface Fermi arcs (SFAs) connect projections of WPs with opposite chirality inside the surface \TB{Brillouin zone} (SBZ)~\cite{Wan:2011}. In contrast, stable \emph{nodal lines} (NLs)~\cite{Burkov:2011,Heikkila:2015b,Bian:2015a,Bian:2015b,Chen:2015,Schoop:2016,Fang:2016,Hirayama:2017} require additional symmetry (such as $\mcP\mcT$ or mirror). They may compose intricate chains~\cite{Bzdusek:2016,Wang:2017,Yu:2017,Gong:2017,Yi:2018}, links~\cite{Yan:2017,Chen:2017,Sun:2017,Chang:2017,Ezawa:2017,Ahn:2018}, knots~\cite{Ren:2017} and nets~\cite{Bzdusek:2016,Feng:2017}, yet the only feature of NL semimetals (NLSMs) on surfaces are usually drumhead states bounded by the projection of bulk NLs~\cite{Heikkila:2011,Delplace:2011,Yu:2015}. Several ways to stabilize \emph{nodal surfaces} were also identified~\cite{Bzdusek:2017,Turker:2017,Zhong:2016,Liang:2016,Tsirkin:2017,Wu:2017}. Various techniques were applied to understand BSNs, including eigenvalue arguments~\cite{Michel:1999,Young:2015,Yang:2015,Zhao:2016b,Yang:2017}, $K$-theory~\cite{Horava:2005,Zhao:2013,Kobayashi:2014,Chiu:2014,Shiozaki:2014,Zhao:2016}, and combinatorics of little-group irreps~\cite{Kim:2015,Po:2017,Kruthoff:2017,Watanabe:2017,Bouhon:2017,Bradlyn:2017}. More recently, \emph{homotopy theory} provided the complete classification of BSNs protected by symmetries local in momentum ($\bk$)-space~\cite{Bzdusek:2017} (i.e. $\mcP\mcT$, $\mcP\mcC$, $\mcC\mcT$ where $\mcP,\mcT,\mcC$ is space-inversion, time-reversal and charge-conjugation symmetry~\cite{Altland:1997}), including nodes carrying \emph{multiple} topological charges~\cite{Fang:2015,Li:2017,Agterberg:2017}. 

In this work, we generalize the homotopy description to systems with additional point-group symmetries. Such a strategy reveals certain features missed by methods~\cite{Michel:1999,Young:2015,Yang:2015,Zhao:2016b,Yang:2017,Horava:2005,Zhao:2013,Kobayashi:2014,Chiu:2014,Shiozaki:2014,Zhao:2016,Kim:2015,Po:2017,Kruthoff:2017,Watanabe:2017,Bouhon:2017,Bradlyn:2017}, as we demonstrate on two-band models with no local-in-$\bk$ symmetries 
%\TB{(class \textrm{A} of Ref.~\cite{Bzdusek:2017})} 
that are either
\begin{itemize}
\item[(\emph{a})] derived from orbitals with opposite eigenvalue of one or more mirror ($m_{x,y,z}$) symmetries, or 
\item[(\emph{b})] symmetric under a $\pi$-rotation ($C_{2z}$) followed by $\mcT$. 
\end{itemize}
A new topological invariant is revealed in both cases, having the following implications: 
\begin{itemize}
\item[(\emph{i})] Mirror-related WPs (which carry opposite chirality) \emph{cannot annihilate} at mirror-invariant planes, but instead \emph{convert} into a NL. The resulting NLSM exhibits \emph{both} SFAs {and} drumhead states. 
\item[(\emph{ii})] Four (eight) WPs related by two (three) mirrors can be fused into two (three) connected NLs (i.e. \emph{nodal chain}) exhibiting multiple SFAs. 
\item[(\emph{iii})] WPs in the presence of $C_{2z}\mcT$ carry \emph{two} integer charges, both of which constrain the conversions of WPs and connectivity of SFAs.
\end{itemize} 
Our conclusions readily apply to several recently investigated systems, such as those of Refs.~\cite{Xu:2011,Xiao:2017,Oh:2017}. \TB{We remark that our analysis applies both to systems with or without spin-orbit coupling (SOC) as long as all local-in-$\bk$ symmetries are broken~\footnote{\TB{In the case of negligible SOC, the spin degree of freedom is dropped from the description. The resulting BSNs are described as ``Weyl-like'' by Ref.~\cite{Chen:2015b}.}}.} \TB{The adiabatic tuning of bulk Hamiltonians considered in our models could be experimentally achieved by the application of strain to the crystalline system.}

We focus on the geometric interpretation of the new invariants, while leaving the formal derivations for the Supplemental Material~\cite{Supp}. Our arguments generalize to systems with more than two bands if the orbitals forming the nodes are sufficiently weakly coupled to other degrees of freedom~\cite{Supp}, although analogous invariants of certain \emph{superconducting} systems are unaffected by the additional bands~\cite{Supp,Fischer:2018}. In a separate work~\cite{Wu:2018}, we apply similar homotopy arguments to analyse nodal chains protected by $\mcP\mcT$, while a general classification scheme will appear elsewhere~\cite{Bzdusek:2018}. 

\emph{Homotopy description}.--- BSNs protected by local-in-$\bk$ symmetries are understood by studying the Hamiltonian $H$ on $p$-dimensional spheres $\sphere{p}$ with spectral gap as follows~\cite{Bzdusek:2017}:
\begin{enumerate} 
\item Local-in-$\bk$ symmetries restrict admissible gapped Hamiltonians $H(\bk)$ to certain \emph{space \TB{of Hamiltonians}} $M$ (assuming the appropriate spectral projection~\cite{Kitaev:2009}).
\item Space $M$ may allow \emph{non-trivial} embeddings of $\sphere{p}$ that {cannot} be continuously deformed to a point (such as $\sphere{1}$ winding around a cylinder). This property is formalized by \emph{homotopy groups} $\pi_p(M)$~\cite{Mermin:1979,Hatcher:2002,Nakahara:2003,Bott:1959,Lundell:1992}.
\item If the image $H(\sphere{p})\!\subset\! M$ of some $\sphere{p} \!\subset\! \textrm{BZ}$ is {non-trivially embedded} in $M$, then $H$ must \emph{be gapless} at some $\bk_\textrm{node}$ inside the \emph{disc} $\disc{p+1}\!\subset\!\textrm{BZ}$ (a contractible region with boundary $\partial\disc{p+1} \!=\! \sphere{p}$). This follows, because \emph{gapped} $H(\disc{p+1})$ admits a continuous contraction of $H(\sphere{p})$ onto a single point $H(\bk_0)\!\in\! M$ for some $\bk_0\!\in\! \disc{p+1}$, thus contradicting the assumed non-trivial embedding. 
\end{enumerate} 
The dimension of the node depends on the lowest non-trivial homotopy group: In three dimensions, $\pi_{0/1/2}$ indicates nodal surfaces/lines/points. For example, consider a two-band model without local-in-$\bk$ symmetries (``class $\textrm{A}$'' of Ref.~\cite{Bzdusek:2017})
\begin{equation}
H(\bk) \!=\! \epsilon(\bk) \b1 + \bd(\bk)\cdot\bsigma \;\;\textrm{with spectrum}\;\; \varepsilon_\pm \!=\! \epsilon(\bk)\pm\abs{\bd(\bk)},\label{eqn:two-band-Ham}
\end{equation}
where $\epsilon$ and $\bd\!=\!(d_x,d_y,d_z)$ are real functions of $\bk$, and $\bsigma\!=\!(\sigma_x,\sigma_y,\sigma_z)$ are the Pauli matrices. The space of gapped Hamiltonians $\mathbb{R}\!\times(\mathbb{R}^3\backslash\{\mathbf{0}\})$ can be projected to $M_\textrm{A} \!\simeq\! \sphere{2}$ by normalizing $\left|{\bd(\bk)}\right|\!=\!1$ and $\epsilon=0$ while preserving the homotopy groups~\cite{Hatcher:2002}. The lowest non-trivial $\pi_2(M_\textrm{A})\!=\!\mathbb{Z}$ suggests nodal points with an integer charge, i.e. WPs.

We now generalize the discussion to systems with point symmetry $R$. Three observations are crucial:
\begin{enumerate}
\item One naturally considers \emph{$R$-symmetric $p$-spheres}. 
\item Since $H(R \bk) = R H(\bk) R^{-1}$, function $H(\sphere{p})$ is fully determined on a \emph{fraction} (the fundamental domain) of $\sphere{p}$.
\item Inside $R$-invariant subspaces $\Pi \!=\! \{\bk_\parallel \!\in\! \textrm{BZ}\,|\, R \bk_\parallel \!=\! \bk_\parallel \}$, 
\begin{equation}
R H(\bk_\parallel) R^{-1} = H(\bk_\parallel), \label{eqn:equator-constraint}
\end{equation}
defining the subspace $X_R \!\subset\! M$ of \emph{$R$-invariant gapped Hamiltonians}. 
\end{enumerate}
For $R\!\in\! \{m_z,C_{2z}\mcT\}$ which flip $k_z \!\mapsto\! -k_z$, the subspaces $\Pi$ are \emph{planes} $k_z\!\in\!\{0,\pi\}$. We set the fundamental domain to be the upper hemisphere, which is topologically a disc $\disc{p}$ with boundary in $\Pi$. We are thus led to study continuous maps $H\!:\! D^p \to M$ with $H(\partial D^p) \subset X_R$, which are classified by \emph{relative homotopy group}~$\pi_p(M,X_R)$~\cite{Hatcher:2002}. For model~(\ref{eqn:two-band-Ham})~\cite{Supp}
\begin{equation}
\pi_2(M_\textrm{A},X_{m}) = \mathbb{Z}\quad\textrm{and}\quad \pi_2(M_\textrm{A},X_{C_{2}\mcT}) = \mathbb{Z}\!\oplus\!\mathbb{Z}\label{eqn:rel-hom}.
\end{equation} 
One can similarly describe systems with more bands and additional symmetries~\cite{Supp,Wu:2018,Bzdusek:2018}. Below, we study the geometric meaning and the physical manifestation of topological charges following from Eqs.~(\ref{eqn:rel-hom}). 
 
\begin{figure}[t!]
\includegraphics[width=0.49 \textwidth]{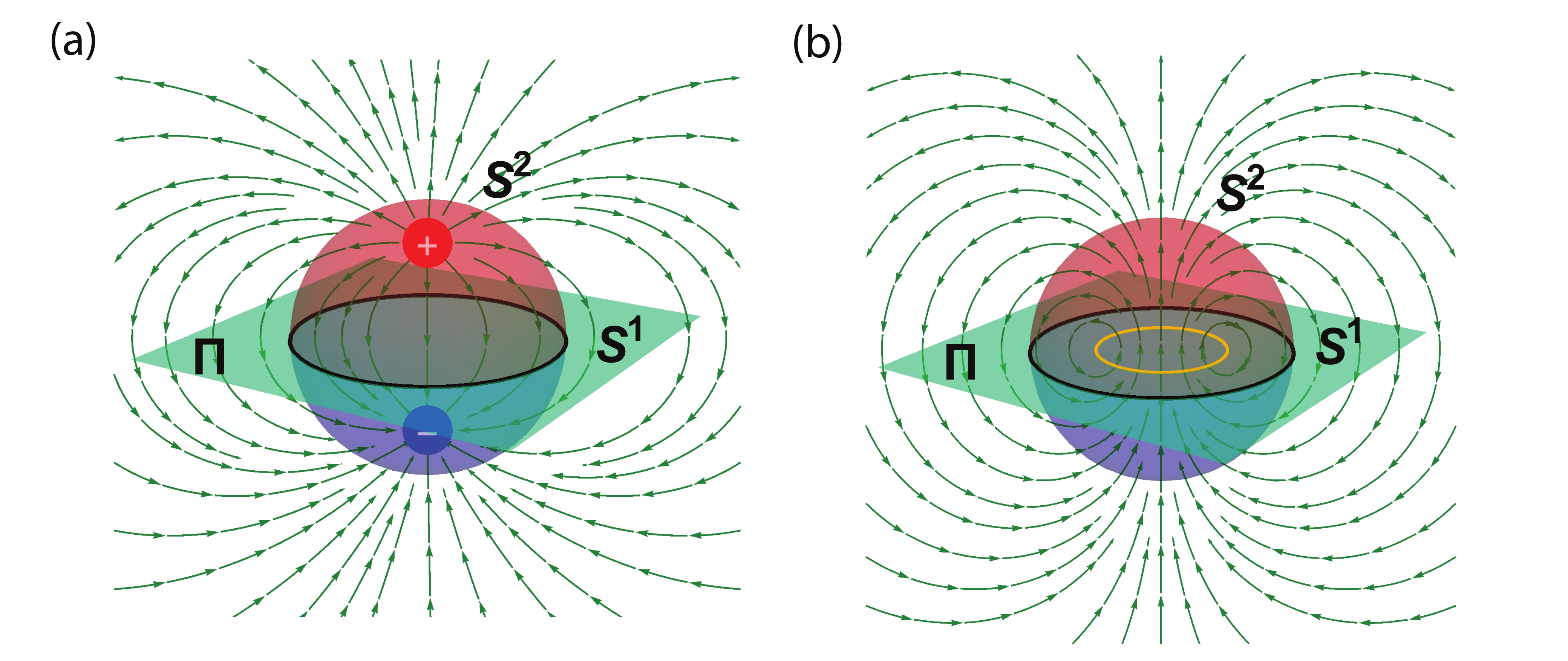}
 \caption{Conversion of mirror ($m_z$)-related Weyl points [red and blue blobs in (a)] into a nodal loop [yellow line in (b)] inside an $m_z$-invariant plane $\Pi$ (cyan). The conversion follows from studying the winding $n_\textrm{W}$ of the pseudospin $\hat{\bd}(\bk)$ (green arrows) on a surrounding sphere $\sphere{2}$ [composed of the upper (red) and the lower (blue) hemisphere, meeting at the equator $\sphere{1}$ (black)]. By the $m_z$-symmetry, $n_\textrm{W}$ has opposite sign on the two hemispheres. Since the winding is the same in both panels, the conversion from (a) to (b) is permitted.}
\label{fig1}
\end{figure}

\emph{Systems with $m_z$ symmetry}.--- We interpret the integer invariant in Eq.~(\ref{eqn:rel-hom}) as the appropriate winding number of the \emph{pseudospin} configuration $\hat{\bd}(\bk)\!=\!\bd(\bk)/|\bd(\bk)|$ of the Hamiltonian in Eq.~(\ref{eqn:two-band-Ham}): For an $m_z$-symmetric $\sphere{2}$ with gapped spectrum, it is sufficient to discuss $H(\bk)$ on a hemisphere on one side of plane $\Pi$~[Fig.~\ref{fig1}]. Since the two basis orbitals \TB{are assumed to} have opposite $m_z$ eigenvalues, we represent $m_z \!\propto\! \sigma_z$, therefore Eq.~(\ref{eqn:equator-constraint}) implies $\hat{\bd}(\bk_\parallel)\!=\!(0,0,\pm 1)$ for $\bk_\parallel$ on the equator $\sphere{2}\!\cap\!\Pi\!\simeq\! \sphere{1}$. The continuity implies that $\hat{\bd}(\sphere{1})$ is a constant, allowing us to identify the equator as one point. This procedure transforms the hemisphere into a \emph{complete} sphere. The integer invariant in Eq.~(\ref{eqn:rel-hom}) accounts for the winding number $n_\textrm{W}$ of $\hat{\bd}(\bk)$ around this sphere. The definition directly generalizes to a generic two-dimensional $m_z$-symmetric manifold with spectral gap. The topological nature implies that $n_\textrm{W}$ cannot change under continuous operations that keep $H$ gapped on the manifold. 

A non-trivial value of $n_\textrm{W}$ on an $\sphere{2}$ indicates BSNs inside the $\sphere{2}$. Let us first examine the situation with two $m_z$-related WPs inside the $\sphere{2}$. One can fully enclose the upper WP by the union of the upper hemisphere with the in-plane disc bounded by the equator [Fig.~\ref{fig1}(a)]. We set the chirality of this WP to $+1$, hence the winding of $\hat{\bd}(\bk)$ on the enclosing manifold is also $+1$. However, since $\hat{\bd}(\bk)$ is constant on the in-plane disc, the winding is completely carried by the hemisphere, implying $n_\textrm{W}\!=\!+1$. Now consider bringing the two WPs together at plane $\Pi$. Such an operation cannot change $n_\textrm{W}$ because it preserves the spectral gap of $H(\sphere{2})$. Importantly, the WPs \emph{cannot annihilate} since this would leave $H(\bk)$ gapped inside the $\sphere{2}$ and thus permit a continuous deformation of the $\sphere{2}$ to a point $\bk_0 \!\in\! \Pi$ with \emph{altered} value $n_\textrm{W} \!=\! 0$. Instead, the WPs must \emph{convert} to an $m_z$-protected NL [described by $d_{z}(\bk_\parallel)\!=\!0$], which is the only generic nodal structure in plane $\Pi$~\footnote{Ref.~\cite{Lim:2017} discussed similar conversions enforced by a more complicated symmetry. Consequently, their NLs do not exhibit drumhead states.}. Conversely, shrinking an in-plane NL with charge $n_\textrm{W}$ leaves behind $\left|n_\textrm{W}\right|$ pairs of residual WPs~\footnote{Without the rotational symmetry, a shrinking NL with $\abs{n_\textrm{W}} = 1$ first splits into a \emph{trivial} NL and a pair of $m_z$-related WPs, before the NL vanishes completely.}. This conversion rule is strict in two-band models with a mirror/glide symmetry, and generalizes to multi-band systems only under the conditions formulated in the Supplemental Material~\cite{Supp}. Note that there is a simple \emph{heuristic} explanation of the conversion based on an analogy with magnetism: Assuming for simplicity $H(\bk)$ with a full rotational symmetry around the $z$-axis, two $m_z$-related WPs create dipolar $\hat{\bd}(\bk)$, resembling the magnetic field of a bar magnet. The dipolar character cannot vanish under continuous deformations of the field. However, the \emph{source} of the dipole can transform into a NL, when $\hat{\bd}(\bk)$ resembles the magnetic field of a loop current [Fig.~\ref{fig1}(b)]. A similar rotationally-symmetric model readily describes BSNs predicted for HgCr$_2$Se$_4$~\cite{Xu:2011}.

\begin{figure}[t!]
\centering
\includegraphics[width=0.49 \textwidth]{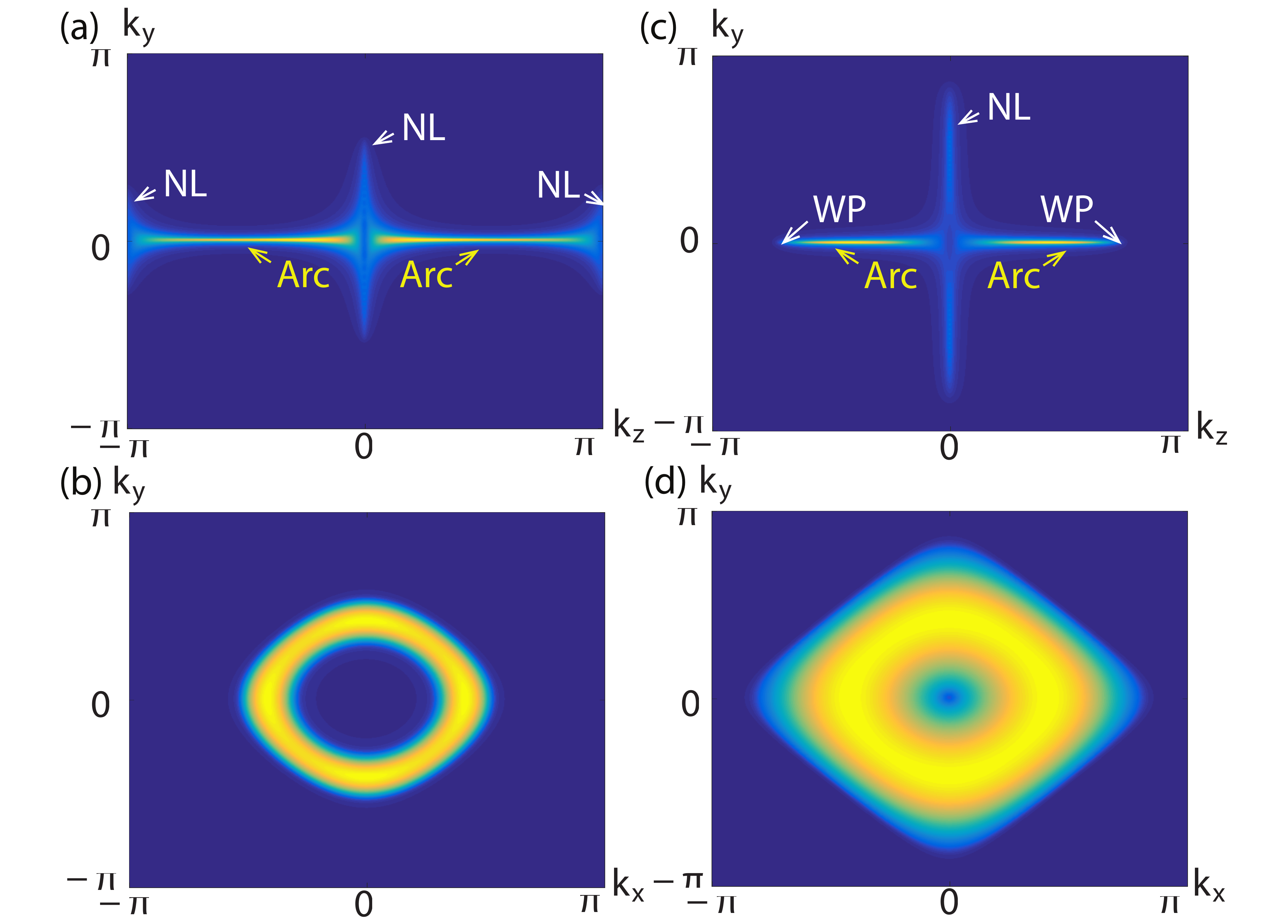}
\caption{Nodal-lines (NLs) exhibiting \emph{both} surface Fermi arcs (SFAs) \emph{and} drumhead states. (a) A model~\cite{Supp} exhibiting a NL with charge $\abs{n_\textrm{W}}\!=\! 1$ in each of the $k_z \!\in\! \{0,\pi\}$ planes. The density of states in the $(100)$ surface Brillouin zone (SBZ) reveals two mirror ($m_z$-)related SFAs, while (b) the $(001)$ SBZ  exhibits drumhead states on an annulus bounded by the projection of the NLs. (c-d) Adjusting the model parameters converts the NL in the $k_z \!=\! \pi$ plane to two Weyl points (WPs). The remaining NL connects with a SFA to both WPs inside the $(100)$ SBZ.}
\label{fig2}
\end{figure}

Topological invariants defined on a 2D manifold inside BZ suggest SFAs passing through the {projection} of the manifold inside SBZ~\cite{Wan:2011}. Since the winding number $n_\textrm{W}$ is protected by $m_z$, one expects such SFAs on $m_z$-symmetric surfaces. Furthermore, since $n_\textrm{W}$ is defined by the winding of $\hat{\bd}(\bk)$ on \emph{either} side of plane $\Pi$, we expect $\left|n_\textrm{W}\right|\!=\!1$ to facilitate a \emph{pair} of $m_z$-related arcs. We test the bulk-boundary correspondence on a model~\cite{Supp} exhibiting a NL with $\abs{n_\textrm{W}}\!=\! 1$ in both of the $k_z\!\in\!\{0,\pi\}$ planes. In the $(100)$ SBZ [Fig.~\ref{fig2}(a)], we indeed observe two SFAs connecting the projections of the NLs. On the other hand, in $(001)$ SBZ [Fig.~\ref{fig2}(b)] we find drumhead states on an annulus bounded by the projection of the two NLs. This is the first model of a NLSM exhibiting \emph{both} robust SFAs \emph{and} drumhead states~\footnote{\TB{Some choices of $\epsilon(\bk)$, $\bd(\bk)$ and of the boundary conditions can detach the SFAs from the NL projections, see~\cite{Supp}.}} on the surface~\footnote{Ref.~\cite{Zhao:2017} considered \emph{real Dirac fermions} with SFAs in the presence of $(\mcP\mcT)^2 = +\mathbf{1}$. However, the $\mathbb{Z}_2$ monopole charge~\cite{Fang:2015} of this symmetry class does \emph{not} in general lead to robust surface states~\cite{Supp}, and fine-tuning is necessary. This is because such SFAs correspond to crossings of counter-propagating surface modes which can hybridize. The SFAs of our models are \emph{chiral}~\cite{Supp} and thus not susceptible to such instabilities.}. A variation of the Hamiltonian parameters~\cite{Supp} converts the NL at $k_z\!=\!\pi$ into two WPs~[Fig.~\ref{fig2}(c-d)]. In this case, the NL at $k_z\!=\!0$ connects with a SFA to both WPs inside the $(100)$ SBZ. An analogous discussion with $\abs{n_\textrm{W}}=2$ explains the surface states predicted for HgCr$_2$Se$_4$ in Ref.~\cite{Xu:2011}. 

\begin{figure}[t!]
\centering
\includegraphics[width=0.465 \textwidth]{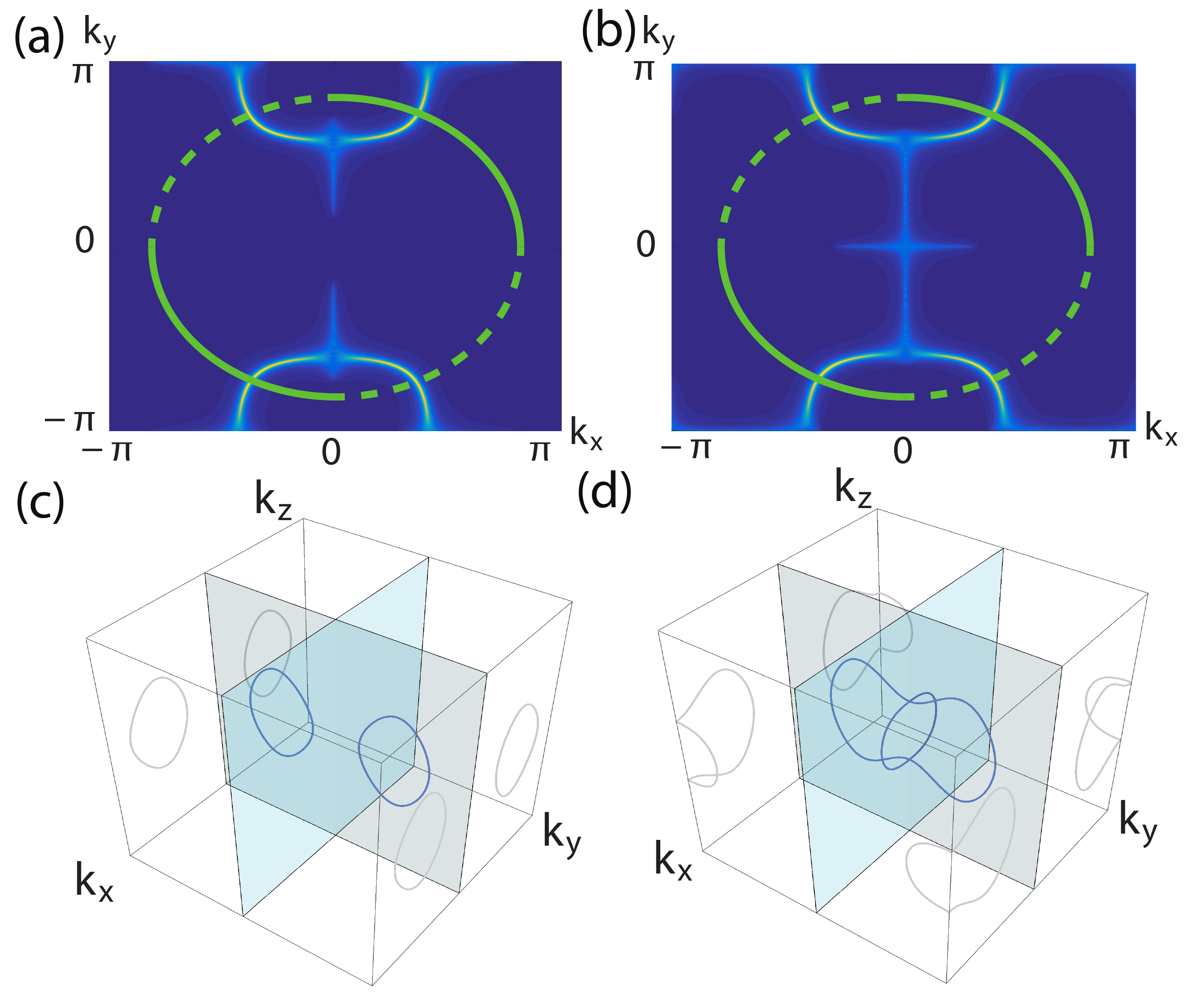}
\caption{Nodal-chain with surface Fermi arcs (SFAa). (a) An $m_{x,y}$-symmetric model~\cite{Supp} exhibiting two nodal lines (NLs) with $\abs{n_\textrm{W}}=1$ in both the $k_{x}=0$ and the $k_{y} = \pi$ plane, each creating the dipolar $\hat{\bd}(\bk)$ of Fig.~\ref{fig1}(b). The winding number on quarter-cylinders projecting onto the green solid/dashed quarter-circles in the (001) SBZ is $n_{\textrm{W}2}=\pm 1$, compatible with the observed connectivity of the SFAs. (b) Adjusting the model parameters produces a nodal-chain phase with four SFAs. (c,d) The nodal structures inside the bulk BZ for parameters producing the surface spectra in panels (a) and (b).}
\label{fig3}
\end{figure}

\emph{Systems with two (three) mirror symmetries}.--- We argued that two mirror-related WPs can coalesce at a mirror-invariant plane and convert into a NL while preserving the dipolar character of $\hat{\bd}(\bk)$. It is natural to consider the quadrupolar (octupolar) generalization facilitated by two (three) perpendicular mirrors. For concreteness, we consider a model based on orbitals differing in \emph{two} mirror eigenvalues (such as $p_{x}$ and $p_{y}$), hence we represent $m_{x,y}\!\propto\!\sigma_z$. Clearly, four $m_{x,y}$-related WPs can pairwise convert into two NLs inside either an $m_x$-invariant or an $m_y$-invariant plane~\footnote{For a schematic summary of the described node conversions, see Fig.~S-6 of~\cite{Supp}.}. Let us assume the first scenario, and study what happens if we \TB{further merge the two resulting} NLs at an $m_y$-invariant plane. From Eq.~(\ref{eqn:equator-constraint}), $H(\bk_\parallel)\!=\!\epsilon(\bk_\parallel) \b1  +  d_z(\bk_\parallel)\sigma_z$ for $\bk_\parallel$ inside both invariant planes. The NLs correspond to boundaries between regions of positive/negative ${d}_z(\bk_\parallel)$. Moving the two NLs together at the $m_y$-invariant plane reverses the sign of $d_z(\bk_\parallel)$ on a segment along an $m_{x,y}$-invariant \emph{line}. This segment connects to finite regions inside \emph{both} invariant planes, thus indicating the presence of \emph{two connected NLs} (a \emph{nodal chain}) [Fig.~\ref{fig3}(c-d)]. Similarly, eight WPs related by three mirrors can fuse into three connected NLs if the basis orbitals differ in \emph{all} mirror eigenvalues (such as $p_z$ and $d_{xy}$).

The topological invariant that enforces the conversion of four WPs to two NLs and further into a nodal chain in the presence of two mirrors follows from Eq.~(\ref{eqn:rel-hom}): Consider an $m_{x,y}$-symmetric $\sphere{2}$ with a spectral gap, and choose the fundamental domain to be the quarter-sphere in the $k_{x,y} \!>\! 0$ quadrant. The boundary of the quarter-sphere carries constant $\hat{\bd}(\bk)$, hence we can identify the boundary with one point. The identification transforms the quarter-sphere into a complete sphere on which we define the winding number $n_{\textrm{W}2}$ of $\hat{\bd}(\bk)$. The winding numbers in neighboring quadrants differ by relative sign, demonstrating the underlying quadrupolar structure. Generalizing the previous discussion, we expect $4\!\left| n_{\textrm{W}2}\right|$ SFAs in the $m_{x,y}$-symmetric $(001)$ SBZ. To demonstrate this feature, we develop a model~\cite{Supp} exhibiting \emph{two} NLs with $\abs{n_\textrm{W}}=1$ in both the $k_{x} = 0$ and the $k_{y}=\pi$ plane [Fig.~\ref{fig3}(c)]. We observe four SFAs connecting the projections of the NLs in the $(001)$ SBZ [Fig.~\ref{fig3}(a)]. The connectivity of the SFAs is compatible with $n_{\textrm{W}2} = \pm 1$ on quarter-cylinders projecting onto the indicated green quarter-circles. Varying the model parameters~\cite{Supp} converts the central two NLs into a nodal chain [Fig.~\ref{fig3}(d)], while preserving $n_{\textrm{W}2}$ and the connectivity of the SFAs [Fig.~\ref{fig3}(b)].

\begin{figure}[b!]
\includegraphics[width=0.49 \textwidth]{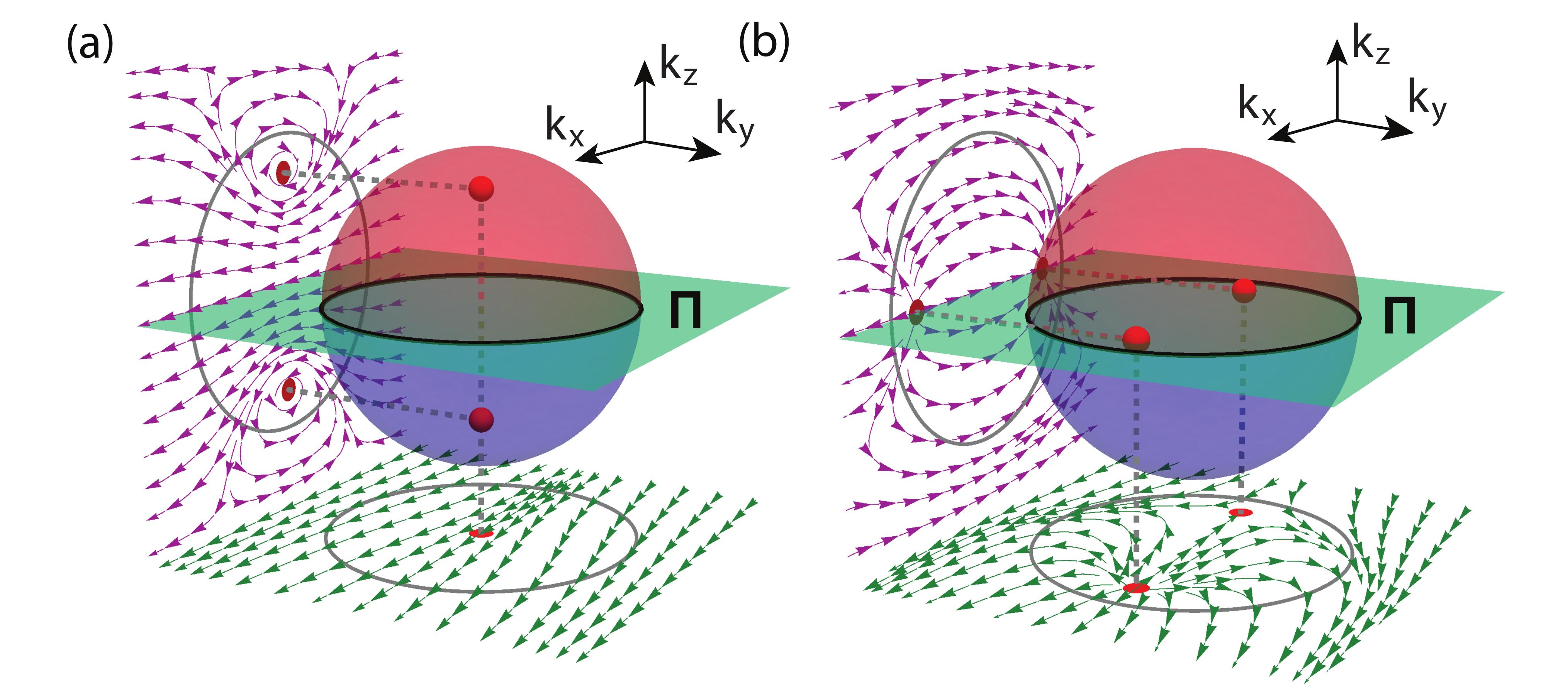}
 \caption{Conversions of Weyl points (WPs, red blobs) in systems with $C_{2z}\mcT$ symmetry conserve \emph{two} integer charges defined on the surrounding sphere [composed of two hemispheres (red and blue) separated at the $k_z\!=\!0$  plane ($\Pi$, cyan) by the equator (black)]. (a) Two out-of-plane WPs related by $C_{2z}\mcT$ carry the same chirality $n_\textrm{C}$, as manifested by the same winding of $\hat{\bd}(\bk)$ in the $k_y \!=\! 0$ plane (magenta projection, left), and induce zero helicity $n_\textrm{H}$ of $\hat{\bd}(\bk)$ in plane $\Pi$ (green projection, bottom). (b) Conversion to in-plane WPs~\cite{Supp} creates helical $\hat{\bd}(\bk)$ in plane $\Pi$. The conservation of $n_\textrm{H}$ requires the obtained WPs to carry \emph{opposite} helicity, as can be checked by following $\hat{\bd}(\bk)$ along the equator projection (grey loop). As $n_\textrm{C}$ is also unchanged, the conversion from (a) to (b) is permitted.}
\label{fig4}
\end{figure}

\emph{Systems with $C_{2z}\mcT$}.--- We finally develop a geometric interpretation of the two integer invariants identified in Eq.~(\ref{eqn:rel-hom}). Since $(C_{2z}\mcT)^2 \!=\! + \b1$ for both \TB{presence/absence of SOC}~\footnote{If $C_{2z}$ is a \emph{screw rotation}, then $(C_{2z}\mcT)^2 \!=\! -\b1$ at $k_z\!=\!\pi$, thus \TB{facilitating} a nodal \emph{surface} on the BZ boundary~\cite{Liang:2016,Tsirkin:2017}.}, we represent $C_{2z}\mcT\!=\!\sigma_{\! x} K$ with the complex conjugation $K$. The relation in Eq.~(\ref{eqn:equator-constraint}) enforces $\bd(\bk_\parallel) \!=\! ( d_x(\bk_\parallel),d_y(\bk_\parallel),0 )$ to be \emph{planar} for $\bk_\parallel \!\in\! \Pi$. Considering a $C_{2z}\mcT$-symmetric $\sphere{2}$ with gapped spectrum, we identify both integer invariants as certain winding numbers of $\hat{\bd}(\bk)$. One invariant corresponds to the winding number $n_\textrm{C}$ of $\hat{\bd}(\bk)$ around the $\sphere{2}$, and counts the total \emph{chirality} of the WPs inside the $\sphere{2}$. The \emph{additional} invariant corresponds to the winding number $n_\textrm{H}$ of the planar $\hat{\bd}(\bk_\parallel)$ along the equator $\sphere{2} \!\cap\! \Pi \simeq \sphere{1}$, and counts the total \emph{helicity} of WPs locked to plane $\Pi$ plane inside the $\sphere{2}$. The locking of WPs to plane $\Pi$ occurs because $C_{2z}\mcT$ \emph{preserves} chirality, implying that a single WP can be the $C_{2z}\mcT$-image of itself~\cite{Wang:2016,Ruan:2016}. WPs lying outside plane $\Pi$ carry $n_\textrm{H}\!=\!0$. Continuous deformations that preserve the spectral gap on $\sphere{2}$ must preserve \emph{both} topological charges $(n_\textrm{C},n_\textrm{H})$. For example, two in-plane WPs can mutually annihilate only if both their charges $(n_\textrm{C},n_\textrm{H})$ cancel out. Alternatively, converting a pair of $C_{2z}\mcT$-related out-of-plane WPs into two in-plane WPs is possible, but the latter must carry opposite helicity. We demonstrate such a scenario explicitly on a model~\cite{Supp} illustrated in Fig.~\ref{fig4}. Even more complicated conversions such as $(\hspace{-0.04cm}+\hspace{-0.04cm},\hspace{-0.04cm}+\hspace{-0.04cm})\to 2\hspace{-0.04cm}\times\hspace{-0.04cm} (\hspace{-0.04cm}+\hspace{-0.04cm},\hspace{-0.04cm}0) \,+\, (\hspace{-0.04cm}-\hspace{-0.04cm},\hspace{-0.04cm}+\hspace{-0.04cm})$ have been observed at $k_z \!=\! 0$ of the model in Ref.~\cite{Xiao:2017}. 

The helicity invariant imposes additional constraints on the connectivity of SFAs inside the $C_{2z}\mcT$-symmetric $(001)$ SBZ: The projection of a cluster of WPs with a total non-zero $n_\textrm{C}$/$n_\textrm{H}$ is expected to connect by a SFA to another WP. The helicity invariant describes a winding pseudospin texture of the states around the bulk WP [Fig.~\ref{fig4}(b)], reminiscent of the helicity of surface Dirac electrons in topological insulators~\cite{Hsieh:2009b}. Assuming that pseudospin $\hat\bd(\bk)$ is correlated with the electron spin, we expect both helicities to be revealed by similar experiments, such as spin-resolved photoemission spectroscopy~\cite{Hsieh:2009,Dil:2009}, circular-polarization dependent photocurrents~\cite{Moore:2010,McIver:2012}, circular dichroism~\cite{Mirhosseini:2012} or spin-transfer torque~\cite{Mellnik:2014,Wang:2015}. For example, we expect a clockwise rotation of a beam of circular polarized light around a surface normal to rotate the induced photocurrent (counter)clockwise for WPs with positive (negative) helicity.

\emph{Conclusions}.--- We generalized the homotopy description of BSNs to systems with additional point-group symmetries. We derived new topological crystalline invariants that enforce non-trivial conversions of WPs and NLs in systems with mirror or $C_{2z}\mcT$ symmetry. We used these invariants to describe a novel NLSM phase exhibiting both Fermi arcs \emph{and} drumhead states on the surface. We demonstrated the conversion rules on simple tight-binding models~\cite{Supp} and investigated the associated surface states. Our results readily apply to certain previously studied systems, such as~HgCr$_2$Se$_4$~\cite{Xu:2011}. The presented analysis can be directly generalized to systems with multiple bands and different choices of symmetry~\cite{Supp,Wu:2018,Bzdusek:2018}.
	
\begin{acknowledgments}
\emph{Acknowledgments}.--- We are grateful to A.~Broido, S.~Fan, M.~H. Fischer, A.~Soluyanov, Q.-S. Wu, M.~Xiao and Q.~Zhou for helpful discussions. X.-Q.S and S.-C.Z. acknowledge support from the US Department of Energy, Office of Basic Energy Sciences under contract DE-AC02-76SF00515. T.B. was supported by the Gordon and Betty Moore Foundation’s EPiQS Initiative, Grant GBMF4302.
\end{acknowledgments}

\bibliography{bibliography}{}
\bibliographystyle{apsrev4-1} 

\end{document}